\documentclass[conference,compsoc]{IEEEtran}
\ifCLASSOPTIONcompsoc
  \usepackage[nocompress]{cite}
\else
  \usepackage{cite}
\fi
\ifCLASSINFOpdf
  \usepackage[pdftex]{graphicx}
\else
\fi
\usepackage{amsmath}
\usepackage{amssymb}
\usepackage{braket}
\begin{document}
%
\title{Thermodynamic Bayesian Inference}

\author{\IEEEauthorblockN{Maxwell Aifer, Samuel Duffield, Kaelan Donatella, Denis Melanson, Phoebe Klett,\\ Zach Belateche, Gavin Crooks, Antonio J Martinez, Patrick J. Coles}
\IEEEauthorblockA{Normal Computing Corporation, New York, New York, USA}}


%


\maketitle

\begin{abstract}
   A fully Bayesian treatment of complicated predictive models (such as deep neural networks) would enable rigorous uncertainty quantification and the automation of higher-level tasks including model selection. However, the intractability of sampling Bayesian posteriors over many parameters inhibits the use of Bayesian methods where they are most needed. Thermodynamic computing has emerged as a paradigm for accelerating operations used in machine learning, such as matrix inversion, and is based on the mapping of Langevin equations to the dynamics of noisy physical systems. Hence, it is natural to consider the implementation of Langevin sampling algorithms on thermodynamic devices. In this work we propose electronic analog devices that sample from Bayesian posteriors by realizing Langevin dynamics physically. Circuit designs are given for sampling the posterior of a Gaussian-Gaussian model and for Bayesian logistic regression, and are validated by simulations. It is shown, under reasonable assumptions, that the Bayesian posteriors for these models can be sampled in time scaling with $\ln(d)$, where $d$ is dimension. For the Gaussian-Gaussian model, the energy cost is shown to scale with $ d \ln(d)$. These results highlight the potential for fast, energy-efficient Bayesian inference using thermodynamic computing.
\end{abstract}


%
\IEEEpeerreviewmaketitle

\section{Introduction}
Bayesian statistics has proved an effective framework for making predictions under uncertainty~\cite{gelman2020bayesian, russo2018tutorial, wilson2020bayesian, duffield2024scalable, abdar2021review, papamarkou2024position}, and it is central to proposals for automating machine learning~\cite{hutter2019automated}.  Bayesian methods enable uncertainty quantification by incorporating prior knowledge and modeling a distribution over the parameters of interest. Popular machine learning methods that employ this approach include Bayesian linear and non-linear regression \cite{bishop2003bayesian}, Kalman filters \cite{sarkka2023bayesian}, Thompson sampling \cite{russo2018tutorial}, continual learning \cite{kirkpatrick2017overcoming, v.2018variational}, and Bayesian neural networks \cite{wilson2020bayesian, duffield2022ensemble}.

Unfortunately, computing the posterior distribution in these settings is often intractable~\cite{izmailov2021bayesian}. Methods such as the Laplace approximation \cite{daxberger2022laplacereduxeffortless} and variational inference \cite{Blei_2017} may be used to approximate the posterior in these cases, however their accuracy struggles for complicated posteriors, such as those of a Bayesian neural network~\cite{izmailov2021bayesian}. Regardless, sampling accurately from such posteriors requires enormous computing resources~\cite{izmailov2021bayesian}. 


Computational bottlenecks in Bayesian inference motivate the need for novel hardware accelerators. Physics-based sampling hardware has been proposed for this purpose, including Ising machines~\cite{inagaki2016coherent, chou2019analog, mohseni2022ising, yamamoto2020coherent, wang2019oim}, probabilistic bit computers~\cite{kaiser2022life,Camsari_2019,aadit2022massively}, and thermodynamic computers~\cite{conte2019thermodynamic, hylton2020thermodynamic, 8123676, coles2023thermodynamic, melanson2023thermodynamic, aifer2023thermodynamic, duffield2023thermodynamic, aifer2024error,
lipka2023thermodynamic,donatella2024thermodynamic}. Continuous-variable hardware is particularly suited to Bayesian inference since continuous distributions are typically used in probabilistic machine learning~\cite{coles2023thermodynamic}. However, a rigorous treatment of how such hardware can perform Bayesian inference with scalable circuits has not yet been given.


The most computationally tractable algorithms for exact Bayesian inference are Monte Carlo sampling algorithms. The Langevin sampling algorithm \cite{neal2012mcmc, welling2011bayesian} is an elegant example inspired by statistical physics, based on the dynamics of a damped system in contact with a heat bath. What we propose in this work is to build a physical realization of the system that is simulated by the Langevin algorithm. The system must be designed to have a potential energy such that the Gibbs distribution $p(x)\propto e^{-\beta U(x)}$ is the desired posterior distribution which is reached at thermodynamic equilibrium. We present circuit schematics for electronic implementations of such devices for Bayesian inference for two special cases. The first is a Gaussian-Gaussian model (where the prior and the likelihood are both multivariate normal, as found in linear regression and Kalman filtering), and the second is logistic regression (where the prior is Gaussian and the likelihood is Bernoulli parameterized by a logistic function). In each case, the parameters of the prior and likelihood are encoded in the values of components of the circuit, and then voltages or currents are measured to sample the random variable.


While thermodynamic algorithms have been proposed for linear algebra~\cite{aifer2023thermodynamic} and neural network training~\cite{donatella2024thermodynamic}, our work can be viewed as the first thermodynamic algorithm for sampling from Bayesian posteriors. Moreover, our work provides the first concrete proposal for non-Gaussian sampling with thermodynamic hardware. Overall, our work opens up a new field of rigorous Bayesian inference with thermodynamic computers and lays the groundwork for scalable CMOS-based chips for probabilistic machine learning.
 

\begin{figure*}[t]
  \centering
  \includegraphics[width=0.95\linewidth]{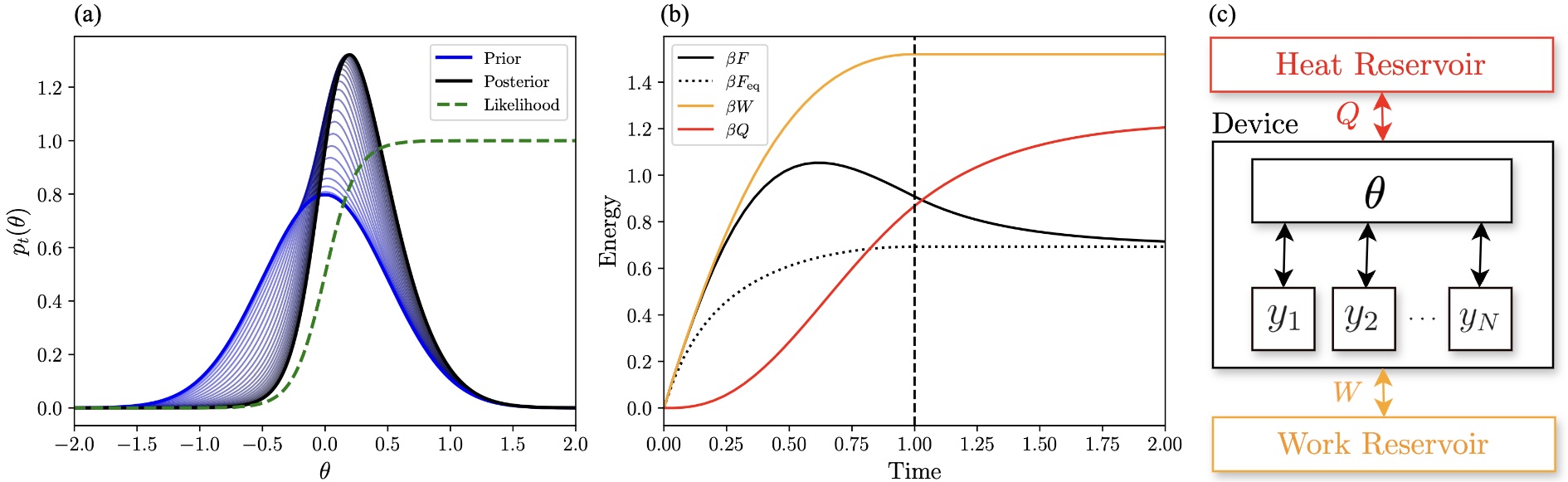}
  \caption{
  \textbf{Overview of Thermodynamic Bayesian Inference.} (a) Time-evolution of the probability density $p_t(\theta)$ under the Fokker-Planck equation for a time-dependent potential energy $U_t(\theta)$. Initially the potential corresponds to the prior $\beta U_0(\theta) = -\ln p_\theta(\theta) $, and a logistic likelihood term is gradually introduced via a quadratic ramp-up until $t=1$, with $U_1(\theta) = -\ln p_\theta(\theta) - \ln p_{y|\theta}(y|\theta)$, after which the potential does not change. The probability density is initially a Gaussian prior (blue) and approaches the Bayesian posterior (black) over time. This is the posterior for logistic regression with a single data point. (b) Thermodynamic quantities during Fokker-Planck evolution. The free energy $\beta F = \beta \braket{U} - S$ (solid black) is computed at all times as well as $F_\text{eq}$ (dotted black), the equilibrium free energy associated with potential $U_t$. Their difference $F - F_\text{eq}$ can be seen as the degree to which the system is out of equilibrium, and equals the KL divergence to the equilibrium distribution. The latter quantity approaching zero at the end of the protocol signifies that $p_t(\theta)$ approaches the true posterior. The convention for the signs of work and heat is $dE = dW - dQ$. The change in free energy over time (dotted black) lower bounds the work (orange), and the gap between the two is the dissipated work. Heat continues to flow after time $t=1$ (the end of the control protocol), while work is done only for $t<1$. (c) Thermodynamic interpretation of Bayesian inference protocol. The parameter $\theta$ and the data points $y_1 \dots y_N$ are physical degrees of freedom, where $y_1\dots y_N$ are fixed and $\theta$ is free to vary. Initially $\theta$ is decoupled from $y_1\dots y_N$, but a coupling is turned on which requires work to be done by a work reservoir. The system exchanges heat with a heat reservoir, coming to thermal equilibrium in the posterior distribution $p_{\theta|y}$.}
  \label{Fig:fig1}
\end{figure*}



We show that in theory the devices proposed for sampling the Gaussian-Gaussian model and logistic regression posteriors can obtain $N$ samples in $d$ dimensions in time scaling with $O(N \ln d)$. This is a significant speedup over typical methods used digitally for the same problems; for example sampling the Gaussian-Gaussian posterior digitally involves matrix inversions taking time scaling with $O(d^\omega)$ where $2 < \omega < 3$. This speedup is larger than the linear (in dimension) speedups found in previous work on thermodynamic algorithms for linear algebra primitives \cite{aifer2023thermodynamic}, where the goal was simply to accelerate standard computations, while not fundamentally changing which problems are considered tractable. In contrast, the more significant speedups found in this work have the potential to make computations possible which previously were not. This is particularly true for the sampling of non-Gaussian posteriors, such as the case of Bayesian logistic regression.


\section{Thermodynamic Bayesian Inference}

Suppose that we have samples of a random vector $y$, and would like to estimate a random vector $\theta$ on which $y$ depends somehow. The Bayesian approach is to assume a prior distribution on $\theta$ given by a density function $p_\theta(\theta)$, and a likelihood function $p_{y|\theta}(y|\theta)$. The posterior distribution for $\theta$ is then given by Bayes's theorem $p_{\theta|y}(\theta|y) = p_{y|\theta}(y|\theta) p_\theta(\theta)/p_y(y)$. To sample from the posterior using the Langevin algorithm, one first computes the score
\begin{equation}\label{eq:score}
    \nabla_\theta \ln p_{\theta|y}(\theta|y) = \nabla_\theta \ln p_{y|\theta}(y|\theta) + \nabla_\theta \ln p_\theta(\theta).
\end{equation}
Then the score is used as the drift term in the following stochastic differential equation (SDE)
\begin{equation}
    \label{eq:langevin-alg-eq}
    d\theta = \nabla_\theta \ln p_{\theta|y}(\theta|y)\, dt_c + \mathcal{N}[0,2\,dt_c],
\end{equation}
where $t_c$ denotes a dimensionless time coordinate. After this SDE is evolved for a sufficient time $T$, the value of $\theta$ will be a sample from $p_{\theta|y}$. This algorithm is equivalent to the equilibration of an overdamped system, as we will now describe. First let $r$ be a vector of the same dimension as $\theta$ describing the state of a physical system, and satisfying
$r = \theta \tilde{r}$
for some constant $\tilde{r}$ (this factor is necessary because $\theta$ is unitless while the physical quantity $r$ has units). Now we define the potential energy function $\beta U(r) = -\ln p_{\theta|y}(r/\tilde{r} \mid y)$. The dynamics of an overdamped system with potential energy $U$ in contact with a heat bath at inverse temperature $\beta$ can be modeled by the overdamped Langevin equation
\begin{equation}
    \label{eq:overdamped-langevin}
    dr = -\gamma^{-1} \nabla_r U(r)\, dt + \mathcal{N}[0, 2\gamma^{-1}\beta^{-1} \, dt],
\end{equation}
where $\gamma$ is a damping constant and $t$ is a physical time coordinate (i.e. having units of time). Note that this implies that $\gamma$ has dimensions of  $\text{energy}\cdot\text{time} / [r]^2$. If we introduce a time constant $\tau = \gamma \beta \tilde{r}^2$, Eq. \eqref{eq:overdamped-langevin} can be written
\begin{equation}
\label{eq:physical-langevin}
   d\theta= \nabla_\theta \ln p_{\theta|y}(\theta|y)\tau^{-1}\,dt + \mathcal{N}[0, 2 \, \tau^{-1}dt],
\end{equation}
which has the same form as Eq. \eqref{eq:langevin-alg-eq}, except with the time constant $\tau$. It is clear that if Eq. \eqref{eq:langevin-alg-eq} must be run for a dimensionless duration $T$ to achieve convergence, then the physical system must be allowed to evolve for a physical time duration $\tau T$ to achieve the same result. While we have addressed the case of conditioning on a single sample $y$ above, the generalization of these ideas to the case of conditioning on multiple I.I.D. samples is given in Appendix~\ref{app_MultipleSamples}. In what follows we will present designs for circuits whose potential energy results in an overdamped Langevin equation that yields samples from Bayesian posteriors.

The process just described is illustrated in Figure \ref{Fig:fig1}, for the example of Bayesian logistic regression with a single data point and a Gaussian prior. In this case $p_\theta(\theta) \propto \frac{1}{2} \sigma^{-2}\theta^2$ and $p_{y|\theta}(y|\theta) = 1/(1+e^{-\theta xy})$, where $y \in \{-1,1\}$ and $x$ is a constant representing a dependent variable. In order to solve a problem using a physical device, we first must map the parameters of the problem onto physical properties of the device, and this is modeled by a control protocol which changes the potential energy over time, specifically\footnote{Note that we abuse notation by writing $U_t(\theta)$ instead of $U_t(r)$ for simplicity of presentation.}
\begin{equation}
    U_t(\theta) = - \ln p_\theta (\theta) - \lambda(t) p_{y|\theta}(y|\theta).
\end{equation}
 The control parameter $\lambda$ is smoothly transitioned from zero to one using a quadratic ramp-up
\begin{equation}
    \lambda(t) = 
\begin{cases}
 - t (t-2) & 0\leq t < 1\\
1 & t \leq 1.
\end{cases}
\end{equation}
The probability density $p_t(\theta)$ evolves according to the Fokker-Planck equation \cite{gardiner1985handbook} (see Appendix~\ref{app_A})
\begin{equation}\label{eq:Fokker-Planck}
    \dot{p}_t= \beta\nabla \cdot (  p_t \nabla U_t) + \nabla^2 p_t,
\end{equation}
with initial condition set by the prior, $p_0 = p_\theta$. The solution to this initial value problem is shown in Fig. \ref{Fig:fig1} (a), and we see the smooth interpolation between the prior (blue) at time $t=0$ and the posterior (black) at time $t=2$.

In theory, the energetic cost of implementing the algorithm is entirely due to the work done during the control protocol for $0\leq t< 1$, as the heat comes from a reservoir which may be taken to be available naturally in the environment.\footnote{In practice, however, the reservoir will be implemented as a random number generator in our proposed devices, and so the power consumed by the random number generators must be included in estimates of energy cost.} The work and heat are quantified using the formulas \cite{seifert2012stochastic}
\begin{equation}
    dW = \int d\theta\, p_t(\theta) d U_t(\theta), \: \: \: dQ = -\int d\theta\, U_t(\theta) dp_t(\theta),
\end{equation}
so the first law $dE = dW - dQ$ is satisfied. The equilibrium free energy $F_\text{eq}$ is defined as the free energy the system would have it were in equilibrium with the potential $U_t(\theta)$, and is given as
\begin{equation}
    \beta F_\text{eq}(t) = -\ln \left( \int d\theta e^{-\beta U_t(\theta)}\right).
\end{equation}
We also may compute the non-equilibrium free energy $\beta F = \beta E - S$. As described in Appendix~\ref{app_A}, the difference between $F$ and $F_\text{eq}$ is simply the KL divergence to the equilibrium distribution \cite{sagawa2020entropy}
\begin{equation}
    \beta (F - F_\text{eq}) = \text{KL}(p\| p_\text{eq}),
\end{equation}
where $p_\text{eq}(\theta) \propto e^{-\beta U(\theta)}$ is the Boltzmann distribution and $\text{KL}(p \| q) = \int d\theta p(\theta) \ln p(\theta)/q(\theta)$.

In Fig. \ref{Fig:fig1} (b), we see that initially $F = F_\text{eq} = 0$, so the system is at equilibrium in the Gaussian prior distribution. However, as the potential changes over time, the system is brought out of equilibrium, and $F$ increases faster than $F_\text{eq}$. For $t \geq 1$ the potential is constant and so the system begins to come back to equilibrium, which can be seen from the fact that $F$ approaches $F_\text{eq}$ for $t \geq 1$. In fact, the approach to equilibrium can be interpreted using the framework of Wasserstein gradient flows (see Appendix~\ref{app_A}). In general, Eq. \eqref{eq:Fokker-Planck} is equivalent to a gradient descent in the space of probability distributions, where the objective function is $F - F_\text{eq}$ and the step size is measured using the Wasserstein 2 metric.

It is important to note the significance of the gap between $F$ and $W$ in Fig. \ref{Fig:fig1}, which represents dissipated work \cite{peliti2021stochastic}. For an idealized device run in the quasistatic limit (ie with a very slowly changing potential), we would have $F = W = F_\text{eq}$ at all times, so the dissipation would be zero and the system would always be at equilibrium. In this case, it would be possible to reverse the protocol and recover an amount of work equal to the work spent during the forward protocol. However, if dissipated work is nonzero, when the protocol is reversed the dissipated work cannot be recovered. Therefore the dissipated work can be seen as the fundamental lower limit on the amount of energy needed to carry out Bayesian inference using this method. This results in a tradeoff between energy and time cost; the dissipated work (and thus the energy cost) can be reduced at the cost of increasing the protocol's time duration.

In Fig \ref{Fig:fig1} (c), the protocol is interpreted as the operation of a kind of thermodynamic machine. The parameter $\theta$ and data $y_1\dots y_N$ are encoded in physical degrees of freedom of a system, with $\theta$ allowed to vary and $y_1 \dots y_N$ fixed. Work is drawn from a reservoir (i.e., a battery) in order to modulate control parameters that couple the $\theta$ subsystem to the subsystems $y_1 \dots y_N$. Heat is exchanged with a heat reservoir, bringing the system towards equilibrium. It follows from the non-negativity of mutual information that the entropy of a Bayesian posterior is (on average) less than or equal to the entropy of the prior, which intuitively means that our certainty increases as we gather more data. For example, in the case of logistic regression, we can see that the entropy of the system decreases as the distribution becomes more sharply peaked (see Fig. \ref{Fig:fig1} (a)). Therefore this machine can be seen as an ``entropy pump", requiring work in order to reduce the entropy of a system while dissipating heat to its environment. Interestingly, it is hypothesized that a similar entropy pumping mechanism is integral to the maintenance of homeostasis in biological systems \cite{svirezhev2000thermodynamics}.

\subsection{Gaussian-Gaussian model}
A particularly simple special case of Bayesian inference is a when both the prior and the likelihood are multivariate normal, and we address this simple model first in order to illustrate our approach more clearly. Specifically, let $\theta\in\mathbb{R}^d$ have prior distribution $ p_\theta(\theta) = \mathcal{N}[\mu_\pi, \Sigma_\pi]$, and let the likelihood be $p_{y|\theta}(y|\theta)=\mathcal{N}[\theta, \Sigma_\ell]$, where $y\in \mathbb{R}^d$ is an observed sample. In this case the posterior $p_{\theta|y}$ is also multivariate normal, with parameters \cite{duffield2022ensemble}
\begin{equation}
    \mu_{\theta|y} = \mu_\pi + \Sigma_\pi \left(\Sigma_\pi + \Sigma_\ell\right)^{-1}(y-\mu_\pi), \label{eq:g-g_postmean}
\end{equation}
\begin{equation}
    \Sigma_{\theta|y} = \Sigma_\pi - \Sigma_\pi (\Sigma_\pi + \Sigma_\ell)^{-1} \Sigma_\pi.  \label{eq:g-g_postcov}
\end{equation}
For this model, the posterior is tractable and can be computed on digital computers relatively efficiently, however for very large dimensions the necessary matrix inversion and matrix-matrix multiplications can still create a costly computational bottleneck. As we will see, the thermodynamic approach provides a means to avoid the costly inversion and matrix products in the computation, and therefore to accelerate Bayesian inference for this model.

\begin{figure}[t]
  \centering
  \includegraphics[width=\linewidth]{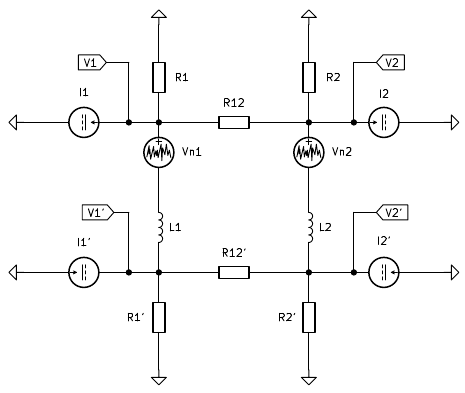}
  \caption{\textbf{Circuit schematic for the Gaussian-Gaussian model posterior sampling device.} Two resistor networks are coupled via inductors, with the currents through the inductors given by \eqref{eq:gaussian-circuit-sde}. }
  \label{Fig:gaussian-circuit}
\end{figure}

We begin by deriving the Langevin equation for sampling this posterior. For this prior and likelihood, the score of the posterior Eq.~\eqref{eq:score} is
\begin{align}
    \nabla_\theta \ln p_{\theta|y}(\theta|y) = -\Sigma_\pi^{-1}(\theta - \mu_\pi) - \Sigma_\ell^{-1}(\theta - y),
\end{align}
and so Eq. \eqref{eq:physical-langevin} becomes
\begin{equation}
\label{eq:gaussian-sde}
    d\theta= -\Sigma_\pi^{-1}(\theta - \mu_\pi)\tau^{-1} dt - \Sigma_\ell^{-1}(\theta - y)\tau^{-1}dt + \mathcal{N}[0, 2 \mathbb{I}\tau^{-1}dt].
\end{equation}
In fact, this SDE can be implemented by a circuit consisting of two resistor networks coupled by inductors, shown in Fig. \ref{Fig:gaussian-circuit} for the two-dimensional case.

\begin{figure}[t]
  \centering
  \includegraphics[width=\linewidth]{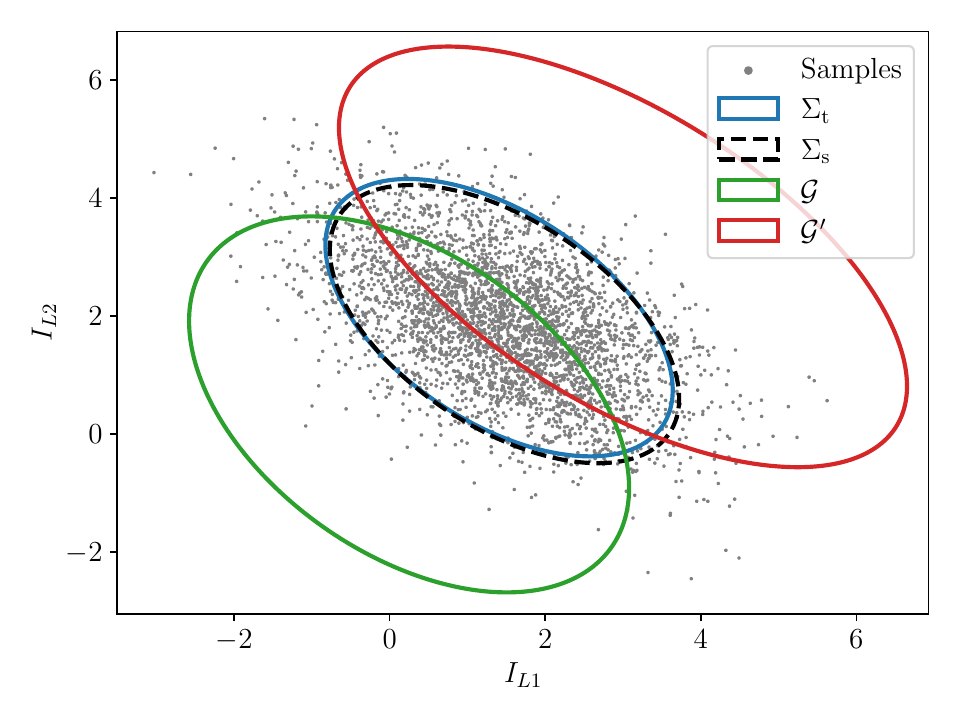}
  \caption{\textbf{SPICE simulations of proposed Gaussian-Gaussian circuit.} The gray points represent the simulated circuit's sampled inductor currents (normalized). The dashed black and solid blue ellipses represent the empirical sample covariance matrix and the target posterior covariance matrix from a Gaussian Bayesian update, respectively. The red and green ellipses represent the prior and likelihood covariance matrices, respectively.}
  \label{Fig:sims_gaussian-circuit}
\end{figure}

The full analysis of the circuit in Fig. \ref{Fig:gaussian-circuit} is given in Appendix~\ref{app_CircuitGG}, but a few remarks are made here to explain its operation. First, we define the conductance matrices $\mathcal{G}$ as
\begin{equation}
\label{eq:conductance_matrix_def}
    \mathcal{G} = \begin{pmatrix}
R_{11}^{-1} + R_{12}^{-1} & - R_{12}^{-1} \\
- R_{12}^{-1} & R_{22} + R_{12}^{-1}
    \end{pmatrix},
\end{equation}
and $\mathcal{G}'$ is defined in the same way for the primed resistors $R_1'$, $R_2'$, and $R_{12}'$. By applying Kirchoff's current law (KCL), the voltages across the resistors can be eliminated. Then the equation $V = L \dot{I}$ is used to derive the following stochastic differential equation for the currents through the inductors
\begin{align}
\label{eq:gaussian-circuit-sde}
    d I_L = &-L^{-1}\mathcal{G}^{-1}(I_L - I)\, dt - L^{-1} \mathcal{G}'^{-1}(I_L - I')\, dt\notag\\
    &+ L^{-1}\sqrt{S} \mathcal{N}[0, \mathbb{I}\, dt],
\end{align}
where $I_L = (I_{L1}\:\: I_{L2})^\intercal$ and $S$ is the power spectral density of each noise source. This equation has the same form as Eq. \eqref{eq:gaussian-sde}, so it is only necessary to determine an appropriate mapping of distributional parameters to physical properties of the circuit's components (see Appendix~\ref{app_CircuitGG}). We verify the behavior of this circuit by running SPICE simulations. The results are shown in Fig.~\ref{Fig:sims_gaussian-circuit} and discussed in greater detail in Section~\ref{sec:experiments}. 

By including more inductors and coupling resistors (as well as current and voltage sources), the design can be generalized to arbitrary dimension.
We note that because of the negative sign in the off-diagonal elements of Eq.~\eqref{eq:conductance_matrix_def}, this specific architecture can only implement matrices with negative off-diagonal elements since passive components cannot achieve a negative conductance. This limitation can be overcome by modifying the architecture and including inductive transformers to change the polarity of the interaction~\cite{melanson2023thermodynamic} or a differential design where symmetry can be exploited to change the direction of the interaction.

The energy and time costs of this algorithm are analyzed in Appendix~\ref{app_TimeEnergyGG} and presented in Section~\ref{sec:complexity}. Numerical simulations are provided in Section~\ref{sec:experiments}.

\subsection{Bayesian linear regression and Kalman filtering}

A generalization of the Gaussian-Gaussian model is that of Bayesian linear regression \cite{bishop2003bayesian} (or equivalently a Kalman filter update step \cite{sarkka2023bayesian, duffield2022ensemble}). In full generality we have
\begin{align}
p_\theta(\theta) &= \mathcal{N}[\mu_\pi, \Sigma_\pi], \label{eq:linear_prior}\\
 p_{y|\theta}(y \mid \theta) &= \mathcal{N}[H\theta, \Sigma_\ell], \label{eq:linear_lik}
\end{align}
Then the overdamped Langevin SDE becomes
\begin{align}
 \label{eq:TLA_SDE}
   d\theta =& - \Sigma_\pi^{-1}(\theta-\mu_\pi)\tau^{-1}\,dt - H^\intercal \Sigma_\ell^{-1} (y - H\theta ) \tau^{-1}\,dt  \notag\\
    &+ \mathcal{N}[0, 2 \mathbb{I} \tau^{-1}dt],\notag \\
= &-(A\theta - b)\tau^{-1}\,   + \mathcal{N}[0, 2 \mathbb{I} \tau^{-1}dt], \\
&\text{  
   for } A = \Sigma_\pi^{-1} + H^\intercal \Sigma_\ell^{-1}H \text{ and } b=\mu_\pi + H^\intercal \Sigma_\ell^{-1}y.\notag 
\end{align}
The form of the SDE (Ornstein-Unhlenbeck process) in \eqref{eq:TLA_SDE} is exactly that of the thermodynamic device in \cite{aifer2023thermodynamic} which if given input $A$ and $b$ above will produce samples from the Gaussian Bayesian posterior $p_{\theta|y}(\theta \mid y)$. Compared to the simpler Gaussian-Gaussian model above, a disadvantage of this approach is that the covariances $\Sigma$ and $\Sigma_\ell$ have to be inverted prior to input as $A$. However, for linear regression, these matrices are often assumed to be diagonal and otherwise they can be efficiently inverted using the thermodynamic procedures in \cite{aifer2023thermodynamic} as preprocessing. Additionally, the formulation of $A$ requires matrix-matrix multiplications which can be costly (even in the case of diagonal covariances). Although, this can be accelerated with parallelization.

On the other hand, the generality of (\ref{eq:linear_prior}-\ref{eq:linear_lik}) makes the approach highly practical. Encompassing Bayesian linear regression \cite{minka2000bayesian} and the update step of the Kalman filter \cite{sarkka2023bayesian}. Moreover in the setting of Kalman filtering, the matrices $\Sigma$ and $\Sigma_\ell$ are typically shared across time points and thus only need to be inverted once in comparison to the Bayesian posterior update which is applied at every time step (and typically represents the computation bottleneck due to the required matrix inversion).


\subsection{Bayesian logistic regression}
Logistic regression is a method for classification tasks (both binary and multiclass) that models the dependence of class probabilities on independent variables using a logistic function. In the Bayesian setting, a prior can be assumed on the parameters of a logistic regression model, for example it is common to assume a Gaussian prior. However, after conditioning on observed data a posterior distribution is produced that has no analytical closed form, making Bayesian logistic regression far less efficient than obtaining a point estimate of the parameters. In this section we present a thermodynamic hardware architecture capable of sampling the posterior for binary logistic regression, and show some preliminary evidence that this architecture can do so more efficiently than existing methods.

Given a parameter vector $\theta\in \mathbb{R}^d$ and an independent variable vector $x\in \mathbb{R}^d$, binary logistic regression outputs a class probability $p_{y|\theta, x}(y|\theta, x)$, where $y\in \{-1,1\}$ (often $y\in\{0,1\}$ is written instead but we choose this notation to simplify the presentation). The likelihood is $p_{y|\theta, x}(y|\theta, x)~=~L(y \theta^\intercal x)$
where $L(z) = 1/(1+e^{-z})$ is the standard logistic function \cite{murphy2022probabilistic}. Note that we will first consider the case of conditioning on a single sample, and in this case the likelihood will be denoted $p_{y|\theta}(y|\theta)$ as $x$ is constant. Additionally, a multivariate normal prior is assumed for the parameters $\theta \sim \mathcal{N}[\mu_\pi,\Sigma_\pi]$. The Langevin equation for sampling the posterior is therefore:
\begin{align}
\label{eq:logistic-langevin-eq}
    d\theta = &-\Sigma_\pi^{-1}(\theta - \mu_\pi)\tau^{-1} dt + L(-y \theta^\intercal x) y x \tau^{-1} dt \notag\\
&+ \mathcal{N}[0,2\mathbb{I}\tau^{-1}dt].
\end{align}
A circuit implementing Eq.~\eqref{eq:logistic-langevin-eq} is shown in  Fig.~\ref{app-logistic-regression-circuit-fig}, and the detailed analysis of this circuit is given in Appendix~\ref{app_CircuitLR}. Equation \eqref{eq:logistic-langevin-eq} is valid for a single data sample, however, as mentioned, in practice we generally take gradients over a larger number of examples such that the gradients are less noisy. This can be done by enlarging the hardware, resulting in the second term of Eq. \eqref{eq:logistic-langevin-eq} being replaced by a sum $\sum_{i=1}^N L(-y_i \theta^\intercal x_i) y_i x_i dt$, with $N$ the number of data points. One may also consider minibatches, and the sum is only over a batch of size $b$. This is achievable by summing currents, which is detailed in the circuit implementation in Appendix~\ref{app_CircuitLR}. At a high-level, implementing this protocol in hardware is very simple in the case of a full batch, since the data only needs to be sent once onto the hardware. The following steps are taken to collect the samples: (1)  Map the data labels to $\{+1, -1\}$. (2) Map the data $(X, Y)$ onto the hardware (full batch setting). (3) Initialize the state of the system, set the mean and the covariance matrix of the prior. (4) At every interval $t_s$ (the sampling time), measure the state of the system $\theta (t)$ to collect samples.

\section{Complexity}\label{sec:complexity}
Compared to previously derived thermodynamic algorithms (such as those in \cite{aifer2023thermodynamic} and \cite{donatella2024thermodynamic}), the algorithms presented here differ in that they do not require the estimation of moments of the equilibrium distribution. For example, in the the algorithm for inverting a matrix \cite{aifer2023thermodynamic}, only a small fraction of the time is spent allowing the system to come to equilibrium, and most of the time is spent collecting samples from this distribution and estimating second moments. Therefore we expect thermodynamic Bayesian inference algorithms can achieve a larger advantage than algorithms based on moment estimation, and we will see that this is indeed the case.

\subsection{Time Complexity}
It has been noted that the concept of time-complexity is somewhat ambiguous for analog computing devices, and generally the energy cost should be accounted for as well \cite{valiant2023matrix}. It is still interesting to consider the physical time necessary to perform an analog computation, and how this scales with the size of the input. In this work, the output is a sample from a probability distribution, and so an appropriate error metric must be used to define the criteria for a successful computation. Here we use the Wasserstein 2 distance between the sampled distribution and the target distribution (see Appendix~\ref{app_A1}), normalized by the norm of the target covariance. 

\begin{figure*}[t]
    \centering
    \includegraphics[width=0.95\linewidth]{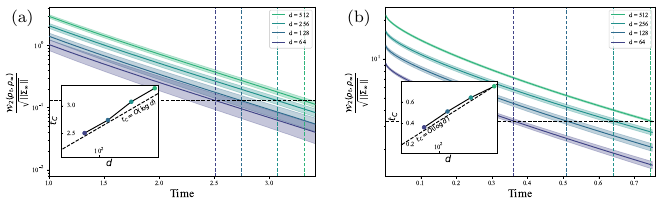}
    \caption{\textbf{Convergence in Wasserstein distance for simulations of Thermodynamic Bayesian Inference.} Here, we consider the Wasserstein distance between simulated thermodynamic samples and the true Gaussian posterior as a function of the number of samples (sampling time). All results are simulated exactly with \texttt{thermox} \cite{duffield2024thermox} and averaged over 50 random seeds with one standard deviation shown. Panel (a): Gaussian-Gaussian model with zero prior mean and covariances sampled from a Wishart distribution. Panel (b): Bayesian linear regression with randomly sampled data and design matrix.}
    \label{fig:numerics-linear}
\end{figure*}

\subsubsection{Gaussian-Gaussian model}
For the Gaussian-Gaussian model, we make the following assumptions
\begin{itemize}
    \item $\|\Sigma_\pi\| \leq 1$ and $\|\Sigma_\ell \|\leq 1$.
    \item $\mu_\pi^\intercal \Sigma_\pi^{-1} \mu_\pi \leq \mathcal{M}_\text{max}$ and $y^\intercal\Sigma_\ell^{-1} y \leq \mathcal{M}_\text{max}$.
\end{itemize}
The first assumption reflects the fact that in order to solve a problem on an analog computing device, the problem must be rescaled to ensure an appropriate signal range for physical dynamical quantities. The second assumption defines how well-conditioned the problem is, and we include the scaling of resources in the parameter $\mathcal{M}_\text{max}$ in our complexity analysis. Subject to these assumptions, it is shown in Appendix~\ref{app_TimeEnergyGG} that in order to have the error bounded as $\mathcal{W}_T^2/\|\Sigma_{\theta|y}\| \leq \varepsilon_W^2$ it is sufficient to allow time $T$ before sampling, given as
\begin{equation}
    T =\tau \ln ((d+2\mathcal{M}_\text{max})\varepsilon_W^{-2}).
\end{equation}
In order to gather $N$ I.I.D. samples from the posterior one simply repeats the protocol $N$ times, resulting in an overall time of $ N\tau \ln ((d+2\mathcal{M}_\text{max})\varepsilon_W^{-2})$. Interestingly, the parameter $\mathcal{M}_\text{max}$ may be allowed to increase linearly with dimension, which preserves the logarithmic scaling in dimension. We thus quote the time complexity for collecting $N$ samples from the posterior as
\begin{equation}\label{eq:gaus-gaus-complexity}
    T = O(N \ln(d \varepsilon^{-2}_W)),
\end{equation}
when $\mathcal{M}_\text{max} \leq O(d)$. It should be noted that this is a worst-case complexity, and the average-case complexity has not yet been fully investigated.

\subsubsection{Logistic Regression}

For our Bayesian logistic regression algorithm, a study of the required energy would require a much more detailed analysis than for the Gaussian-Gaussian model, and is outside the scope of this work. However, it is possible to derive bounds on the required time with less effort, which is done in Appendix~\ref{app_TimeLR}. In particular, assuming $\|\Sigma_\pi\| \leq 1$, we find that it suffices to allow time $T$ before sampling, given as
\begin{equation}
T = \tau \ln((d + \mathcal{M}_{\theta|y})\varepsilon^{-2}),
\end{equation}
where $\mathcal{M}_{\theta|y} = \mu_{\theta|y}^\intercal\Sigma_{\theta|y}\mu_{\theta|y}$. This result leaves something to be desired, as it involves the posterior mean and covariance, and as of yet we have no results constraining the scaling of these parameters with dimension. However, if we introduce an ad-hoc assumption that $\mathcal{M}_{\theta|y}\leq O(d)$ then we obtain
\begin{equation}
\label{eq:time-complexity-LR}
    T = O(N \ln (d  \varepsilon^{-2})).
\end{equation}

\subsection{Energy Complexity}
Some skepticism is warranted of treatments that provide the scaling of time required for analog computation alone. This is because the analog device itself will have to grow with dimension, possibly affording some additional parallelism; in other words, it is possible that dimensional scalings appear too favorable because the computational resources available also grow with dimension. It is therefore essential to also investigate the scaling of energy with dimension, which allows for a fairer comparison to other computational paradigms where the computational resources do not grow with problem size.

\begin{figure*}[t]
    \centering
    \includegraphics[width=0.9\linewidth]{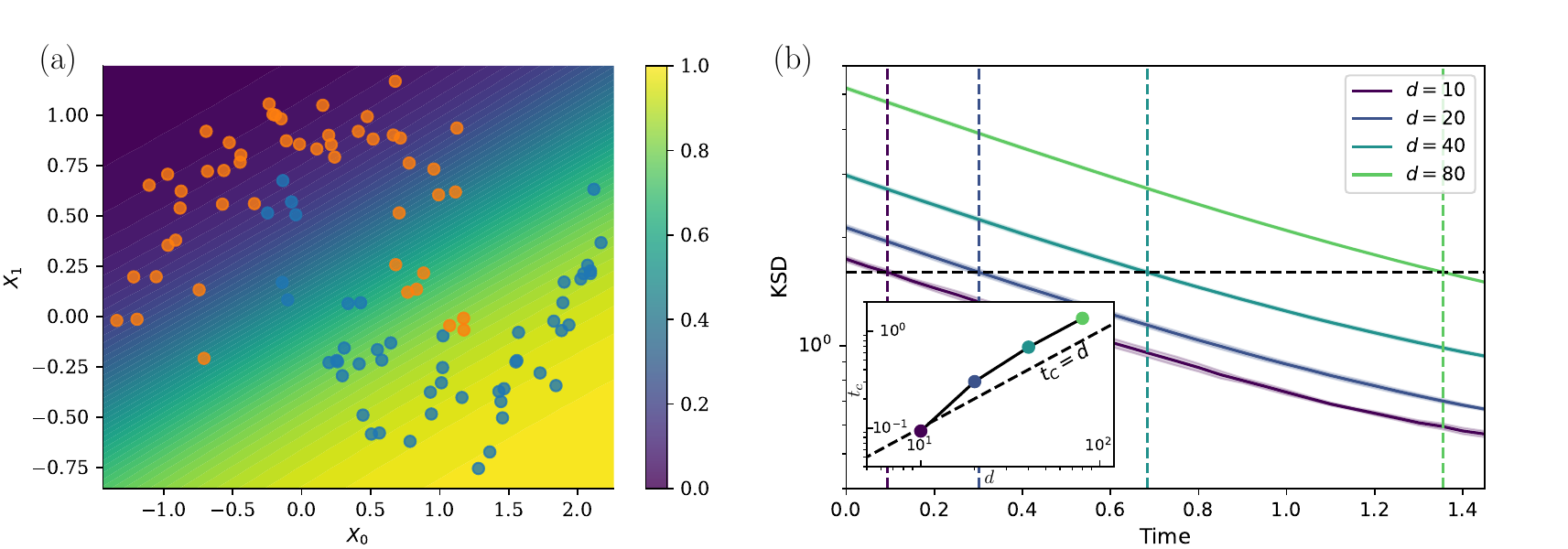}
    \caption{\textbf{Bayesian logistic regression on a two-moons dataset.} Panel (a): Contour plot of the probability for a point in the ($x_1, x_0$) plane to belong to class 1 (blue points). The dataset is also shown, where class 0 (blue points) and class 2 (orange points) are arranged in two intersecting moons. This corresponds to a logistic regression problem with $d=2$ features. Panel (b): Kernel Stein discrepancy (KSD) of samples as a function of time with an ideal thermodynamic sampler for varying dimensions. Inset: extracted scaling of the crossing time as a function of dimension for a chosen crossing value of the KSD, similarly to other experiments. The color of each point matches the color of vertical dashed lines in the main figure which indicates the crossing time for each considered dimension. The KSD is averaged over five different runs, and for each time $1000$ samples are collected.}
    \label{fig:numerics-logistic-regression}
\end{figure*}

The energy required for the Gaussian-Gaussian protocol is derived in Appendix~\ref{app_TimeEnergyGG}, where we use the first law of thermodynamics $\Delta E = W - Q$ to compute the work done by the voltage and current sources in the circuit. In this case, the change in internal energy is associated with the inductors and the dissipated heat with the resistors. The same assumptions are made as were used for the time analysis, leading to the following expression for a sufficient amount of work
\begin{equation}
    W = 2\tilde{I}^2 L\mathcal{M}_\text{max} \ln((d+ 2\mathcal{M}_\text{max})\varepsilon_W^{-2}) + \frac{1}{2}\tilde{I}^2 L(d+ 2\mathcal{M}_\text{max}).
\end{equation}
Once again, this result is simply multiplied by $N$ in order to collect $N$ I.I.D. samples. Assuming, as before, that $\mathcal{M}_\text{max} \leq O(d)$, the following scaling is found
\begin{equation}
    E = O(N d \ln (d \varepsilon^{-2}_W)).
\end{equation}
Once again, this is a worst-case result, and the treatment of the average case remains for future work.

\section{Experiments}\label{sec:experiments}

\subsection{Gaussian-Gaussian model}

To verify that the proposed circuit in Fig.~\ref{Fig:gaussian-circuit} does indeed evolve according to the correct SDE, we ran SPICE circuit simulations. Figure~\ref{Fig:sims_gaussian-circuit} shows the results of such a simulation where a 2-dimensional Gaussian prior and a 2-dimensional Gaussian likelihood are encoded into the conductances while the current in each inductor is measured to determine the resulting posterior. The circuit simulations show strong agreement between the theoretical prediction of the posterior and the simulated current distribution.

The system is simulated for 100 $\mu$s with a sampling rate of 2.0 MHz and a burn-in period of 10 $\mu$s. For clarity, only a small portion of the 180 000 simulated current samples used for the empirical covariance calculation are plotted in Fig.~\ref{Fig:sims_gaussian-circuit}. The conductance matrices are
\begin{equation}
    \mathcal{G} = \begin{pmatrix}
        2.0 & - 1.0 \\
        - 1.0 & 2.5
    \end{pmatrix}
    \mathrm{and}~
    \mathcal{G}' = \begin{pmatrix}
        3.3 & - 2.0 \\
        - 2.0 & 3.2
    \end{pmatrix}.
\end{equation}
The current sources are set to $I = \left(0.3, 0.5\right)^\intercal$ and $I' = \left(3.0, 3.0\right)^\intercal$. Finally, the scales involved are as follows: $\tilde{I} = 1.0~\mathrm{mA}$, $\tilde{R} = 1.0~\mathrm{k}\Omega$, and $L = 1.0~\mu\mathrm{H}$.

In Figure~\ref{fig:numerics-linear}(a), we report the convergence of simulated thermodynamic samples for the Gaussian-Gaussian model with zero prior mean and covariances $\Sigma, \Sigma_\ell$ randomly sampled from a Wishart distribution with $2d$ degrees of freedom. We see fast $O(\log(d))$ convergence in Wasserstein distance between $p(\theta_t \mid \theta_0)$ and the true posterior, supporting our theoretical claims \eqref{eq:gaus-gaus-complexity}.

\subsection{Bayesian linear regression}

In Figure~\ref{fig:numerics-linear}(b), we simulate the evaluation of the thermodynamic linear algebra device \cite{aifer2023thermodynamic} for a Bayesian linear regression task. We sample synthetic data $y \in \mathbb{R}^n$ from $p(\theta) = \mathcal{N}(0, \mathbb{I})$ and $p(y \mid \theta) = \mathcal{N}(n^{-\frac12}H\theta, \mathbb{I})$ with random elements of the design matrix $[H]_{ij} \sim \mathcal{N}(0,1)$ and the $n^{-\frac12}$ scaling ensures data $y$ is normalized (a standard practice in machine learning). We fix $n=500$ and vary the dimension of $\theta$. We observe that the Wasserstein distance to the posterior converges rapidly, matching the logarithmic convergence in the numerics of the Gaussian-Gaussian model.

\subsection{Bayesian logistic regression}

In Fig.~\ref{fig:numerics-logistic-regression}(a), we present results for a Bayesian logistic regression on a two-moons dataset, made of points separated in two classes that are arranged in intersecting moons in the 2D planes, as shown in Fig.~\ref{fig:numerics-logistic-regression}(a). These results are obtained by running the SDE of Eq.~\eqref{eq:logistic-langevin-eq} for $d=2$, hence corresponds to an ideal simulation of the thermodynamic hardware. In this scenario, there are 3 parameters to sample, and $N = 100$ points were considered. 
In Fig.~\ref{fig:numerics-logistic-regression}(a), we see that even for such a simple model, only a few points are misclassified. As mentioned, previously, this setting also gives access to better methods to estimate uncertainty in predictions. 

In Fig.~\ref{fig:numerics-logistic-regression}(b), the Kernel Stein discrepancy (KSD)~\cite{gorham2015measuring} of samples as a function of time is shown for varying dimensions. For these experiments, a dataset made of random points in the $d$-dimensional hyperplane were generated randomly, belonging to two classes as in the two-moons dataset. The figure displays an exponential scaling of the KSD towards its final value (which is not necessarily zero, see~\cite{gorham2015measuring}), similarly to the Wasserstein distance in other experiments. The inset of Fig.~\ref{fig:numerics-logistic-regression}(b) shows the extracted scaling of the crossing time $t_c$ as a function of dimension, which was obtained by fixing a KSD threshold and extracting after how much time this threshold was reached for various dimensions. While our Eq. \eqref{eq:time-complexity-LR} predicts a logarithmic scaling of time with dimension for this problem, we observe an approximately linear scaling of the time to reach a given KSD. This discrepancy could be explained by a number of factors; for one, it is not clear that our simulations go to high enough dimension to reveal asymptotic behavior. Also, due to the intractability of the true posterior we use the KSD rather than Wasserstein 2 distance. The KSD requires a kernel specification with a bandwidth that we keep constant across dimension, which may influence dimensional scaling. Finally, unlike our simulations for Gaussian sampling, this system cannot be simulated exactly, so it is possible that there is a dimension-dependent bias due to time discretization (i.e., finite step size).

%

\section{Conclusion}

The connection between Bayesian inference and thermodynamics has been highlighted previously~\cite{lamont2019correspondence,FRISTON200670,still2012thermodynamics,bartolotta2016bayesian}, although largely in an abstract sense. In this work, we proposed a concrete approach to sampling Bayesian posteriors based on thermodynamics. In Thermodynamic Bayesian Inference (TBI), observed data is encoded in constraints on a physical system, whose degrees of freedom represent the variables we would like to learn about. The process of learning from data is accomplished via the natural equilibration of the system, or equivalently the minimization of its free energy with respect to the imposed constraints. The device that accomplishes this can be viewed as an ``entropy pump", which requires work to be done in order to reduce the entropy of a system while emitting heat to its environment. Interestingly, it has been put forward that similar mechanisms are used in biological systems (in particular, the brain) for maintaining homeostasis and learning from experience~\cite{FRISTON200670, svirezhev2000thermodynamics}.

Beyond these conceptual insights, our work has direct practical relevance. We provided explicit constructions of CMOS-compatible analog circuits to implement our TBI algorithms with scalable silicon chips. Our circuit for performing logistic regression represents the first concrete proposal for non-Gaussian sampling with a thermodynamic computer. It is widely acknowledged that non-Gaussian sampling is difficult for digital computers, and often avoided digitally by imposing Gaussian approximations. Thus, our non-Gaussian sampling approach could open up qualitatively new algorithms that otherwise would be avoided due to their difficulty.

In the cases of Gaussian Bayesian inference (Gaussian prior, Gaussian likelihood) and logistic regression, our analysis showed a sublinear complexity in $d$, leading to a speedup over standard digital methods that is greater than linear. This is an even larger speedup than those previously observed for thermodynamic linear algebra~\cite{aifer2023thermodynamic}, suggesting that Bayesian inference is an ideal application for thermodynamic computers. Our work lays the foundation for accelerating Bayesian inference, a key component of probabilistic machine learning, with physics-based hardware.

Given that the use of thermodynamic computing for Bayesian inference has not been previously explored, many open questions remain. Immediate extensions of our results include designing a circuit realization of our algorithm for Bayesian linear regression, and quantifying the energy consumption of our Bayesian logistic regression protocol. A farther reaching goal is to understand the fundamental limits on Bayesian inference imposed by thermodynamics, in terms of resources including energy and time. In general, we believe that the impact of this work is not only in providing a new means of implementing Bayesian computations, but additionally our results yield a new perspective on Bayesian inference through the lens of thermodynamics.

\ifCLASSOPTIONcompsoc




%

\bibliographystyle{unsrt}
\bibliography{tbi.bib}
\clearpage

\onecolumn
\appendices

\section{Preliminaries}\label{app_A}
Here we introduce some mathematical concepts that will be useful for understanding the derivations in the Appendix.




\subsection{Wasserstein distance}\label{app_A1}

When analyzing the time evolution of probability distributions with Langevin dynamics, the Wasserstein distance plays a key role. While it actually a continuous family of distances with hyperparameter $p$, here we focus on the Wasserstein distance with $p=2$, which can be defined as:
\begin{equation}
    \mathcal{W}_2 (\rho_1, \rho_2) = \inf (\mathbb{E}|X-Y|^2)^{1/2}
\end{equation}
where the infimum is taken over all random variables $X$ and $Y$ whose respective probability densities are $\rho_1$ and $\rho_2$. We will often omit the subscript 2 (and use the subscript to denote something else), as in this work we always use $p=2$. This distance invites alternative interpretations in terms of optimal transport or fluid mechanics. In the latter case, one can write 
\begin{equation}
    \mathcal{W}_2 (\rho_1, \rho_2) = \min_{v} \int_{0}^1 \int_{\mathbb{R}^n} \| v(x,t)\|^2\rho(x,t)dxdt
\end{equation}
where the density $\rho$ and velocity $v$ satisfy a continuity equation: $\dot{\rho}+\nabla \cdot (\rho v)=0$ with boundary conditions: $\rho(t=0) = \rho_1$ and $\rho(t=1) = \rho_2$.

Below we discuss how the Wasserstein distances enables one to reinterpret Langevin dynamics as a gradient descent process.


\subsection{Wasserstein gradient flows}

There is an elegant reinterpretation of Langevin dynamics in terms of gradient descent, which is often called Wasserstein gradient flows~\cite{jordan1998variational}. In this formulation, we consider a recursive stepping rule, where the probability density $\rho^{(k-1)}$ is updated to a new density $\rho^{(k)}$ given by
\begin{equation}\label{eqn_WGF}
    \rho^{(k)} = \min_{\rho \in K} \left(\frac{1}{2} \mathcal{W}_2 (\rho^{(k-1)}, \rho)^2 + h F(\rho)\right).
\end{equation}
Here, $K$ is the set of all probability densities with finite second moment and $h$ can be viewed as a step size (analogous to the learning rate in gradient descent). The function 
\begin{equation}
   F(\rho) = E(\rho) - S(\rho)/\beta
\end{equation}
is the free energy, where the energy and entropy functions are given by
\begin{equation}
    E(\rho) = \int_{\mathbb{R}^n} U(x)\rho(x) dx\,,\qquad S(\rho) = - \int_{\mathbb{R}^n} \rho(x)\log \rho(x) dx.
\end{equation}
The key connection is that the recursive evolution of $\rho^{(k)}$ given by Eq.~\eqref{eqn_WGF} exactly matches the time evolution of $\rho$ given by the following Fokker-Planck equation:
\begin{equation}\label{eqn_FPE}
    \dot{\rho} = \nabla \cdot (\nabla U(x) \rho) + (1/\beta) \nabla^2 \rho,
\end{equation}
where $U(x)$ is identified as the potential energy function and $\beta$ as the inverse temperature. Of course, this Fokker-Planck equation can be alternatively written in terms of its associated Langevin dynamics:
\begin{equation}
    d X_t =  - \nabla U(X_t) dt + \sqrt{2 /\beta} dW_t
\end{equation}
where $W_t$ is the standard Weiner process, and $X_t$ is a random variable whose probability density is $\rho$. Thus, Ref.~\cite{jordan1998variational} found a deep connection between Langevin dynamics and the recursive process in \eqref{eqn_WGF}, which can be viewed as a gradient descent process. Specifically, \eqref{eqn_WGF} involves gradient descent on the free energy function $F(\rho)$, in a space where the Wasserstein distance is the metric of choice (hence the name Wasserstein gradient flow). This shows that Langevin dynamics attempt to minimize the free energy function.

\subsection{Connection between free energy and KL divergence}

We remark that there is a close connection between the free energy function and the Kullback-Leibler (KL) divergence. Specifically, suppose that we consider the normalized stationary state of the Fokker-Planck equation, $\rho_{\infty}(x) =Z^{-1} e^{-\beta U(x)}$, where $Z = \int dx e^{-\beta U(x)}$. Then the KL divergence to this state is given by:
\begin{align}
    \text{KL}(\rho || \rho_{\infty}) &= \int_{\mathbb{R}^n}dx\, \rho(x) \ln \frac{\rho(x)}{\rho_{\infty}(x)}dx \\
&= - S[\rho] -  \int_{\mathbb{R}^n} dx\,  \rho(x)(-\beta U(x) - \ln(Z)) dx\\
& = \beta F[\rho] + \ln(Z)\\
& = \beta (F[\rho] - F[\rho_\infty]),
\end{align}
where in the last line we used the equilibrium thermodynamic identity $\beta F = -\ln(Z)$. So we see that the KL divergence to the stationary state is proportional to the free energy. 
In light of this connection, we can reinterpret the Wasserstein gradient flow from Eq.~\eqref{eqn_WGF} as gradient descent on the KL divergence to the stationary state. Therefore, time evolution under the Fokker-Planck equation in \eqref{eqn_FPE} progressively minimizes this KL divergence. 

\subsection{Convergence to the stationary distribution}

Under certain conditions, any initial probability density $\rho_0$ will evolve over time under Langevin dynamics such that it approaches the stationary state $\rho_{\infty}$ exponentially in time. One can capture this notion quantitatively, for example, with either the KL divergence or the Wasserstein distance.

With the KL divergence, suppose we assume that the stationary state $\rho_{\infty}$ satisfies the log-Solobev inequality (LSI) with constant $1/\alpha$. Then it follows that the state at time $t$, $\rho_t$, satisfies:
\begin{equation}\label{eqn_KL_convergence}
   \text{KL}(\rho_t || \rho_{\infty})\leq e^{-2\alpha t} \text{KL}(\rho_0 || \rho_{\infty}).
\end{equation}
This implies that the KL divergence to $\rho_{\infty}$ shrinks monotonically with time $t$, since we can choose $\rho_0$ to be $\rho_{t - \Delta t}$ where $\Delta t$ is small shift in time. The LSI condition that is required for \eqref{eqn_KL_convergence} is an example of isoperimetry, see for example Ref.~\cite{vempala2019rapid} for a discussion of isoperimetry. 
A simpler condition that is sufficient for convergence is positive curvature of the potential\cite{bolley2012convergence}. Specifically, suppose it holds that 
\begin{equation}
   \nabla^2 U(x) \geq (\alpha/\beta) \mathbb{I},
\end{equation}
where $\nabla^2$ denotes the Hessian, $\mathbb{I}$ is the identity matrix and the stationary state of the Fokker-Planck equation is $\rho_{\infty}(x) \propto e^{-\beta U(x)}$. Then as shown in Ref.~\cite{bolley2012convergence}, it follows that the ($p=2$) Wasserstein distance contracts exponentially in time:
\begin{equation}
   \mathcal{W}_2(\rho_t,\sigma_t)\leq e^{-\alpha t} \mathcal{W}_2(\rho_0,\sigma_0),
\end{equation}
for any two initial states $\rho_0$ and $\sigma_0$, and their corresponding states at time $t$, $\rho_t$ and $\sigma_t$. Because we can always choose $\sigma_0$ to be the stationary state, it follows that the Wasserstein distance to the stationary state shrinks exponentially in time:
\begin{equation}
   \mathcal{W}_2(\rho_t,\rho_{\infty})\leq e^{-\alpha t} \mathcal{W}_2(\rho_0,\rho_{\infty}).
\end{equation} 
Thus, positive curvature guarantees exponential convergence in time.



\subsection{First law of thermodynamics}

The above preliminaries are useful when analyzing the time dynamics (i.e., the runtime cost) of thermodynamic algorithms. However, we are also interested in analyzing the energy dynamics (i.e., the energy cost) of such algorithms.

For the purpose of analyzing energy dynamics, it is useful to review the first law of thermodynamics. This law is essentially a statement of the law of conservation of energy. Whenever no matter is exchanged between a system and its environment, the change in internal energy $\Delta E$ of the system is the work $W$ done on the system and minus the heat $Q$ that is lost (i.e., dissipated) from the system:
\begin{equation}\label{eqn_thermo1stLaw}
   \Delta E = W - Q.
\end{equation} 
Below we will employ this law when analyzing the energy cost of our thermodynamic algorithms (see Appendix~\ref{app_TimeEnergyGG}).

\section{Analysis of Gaussian Bayesian Inference Circuit}\label{app_CircuitGG}

\begin{figure}[h]
  \centering
  \includegraphics[width=0.5\linewidth]{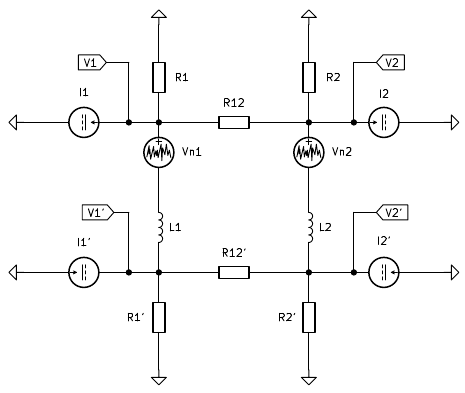}
  \caption{\textbf{Circuit schematic for the Gaussian-Gaussian model posterior sampling device.} Note: This figure has the same content as Fig.~\ref{Fig:gaussian-circuit}, but is repeated here for convenience.}
  \label{app:gaussian-circuit-fig}
\end{figure}

In Figure \ref{app:gaussian-circuit-fig}, positive current goes up through the two inductors, left to right through $R_{12}$ and $R_{12}'$, and towards ground in the other resistors. The two inductors have the same inductance $L$. KCL gives
\begin{equation}
    I_{L1}-I_1 = I_{R1}+I_{12}
\end{equation}
\begin{equation}
    I_{L2}-I_2=I_{R2}-I_{12}
\end{equation}
\begin{equation}
    -I_{L1}+I_{1}'=I_{R1}'+I_{12}'
\end{equation}
\begin{equation}
    -I_{L2}+I_{2}'=I_{R2}'-I_{12}'.
\end{equation}
Using Ohm's law,
\begin{equation}
I_{L1}-I_1 = R_1^{-1}V_1+R_{12}^{-1}(V_1-V_2)=(R_1^{-1}+R_{12}^{-1})V_1 - R_{12}^{-1}V_2
\end{equation}
\begin{equation}
    I_{L2}-I_2=R_2^{-1}V_2-R_{12}^{-1}(V_1-V_2)=(R_2^{-1}+R_{12}^{-1})V_2-R_{12}^{-1}V_1.
\end{equation}
These can be written as a single vector equation as follows
\begin{equation}
    I_{L}-I = \mathcal{G}V,
\end{equation}
where $I_L = (I_{L1} \: \: I_{L2})^\intercal$,  $I = (I_1 \: \: I_2)^\intercal$, and
\begin{equation}
    \mathcal{G} = \begin{pmatrix}
R_1^{-1}+R_{12}^{-1} & - R_{12}^{-1}\\
-R_{12}^{-1} & R_2^{-1} + R_{12}^{-1}
\end{pmatrix}.
\end{equation}
Similarly, for the lower subcircuit we have
\begin{equation}
    -I_{L}+I'=\mathcal{G}'V'.
\end{equation}
The inductors obey the equations
\begin{equation}
    L_1\dot{I}_{L1} = V_1'-(V_1-V_{n1})
\end{equation}
\begin{equation}
L_2\dot{I}_{L2}=V_2'-(V_2-V_{n2}),
\end{equation}
or in vector notation
\begin{equation}
    L \dot{I}_L = V' - V + V_n.
\end{equation}
Substituting in the expressions for $V$ and $V'$ derived before, we have
\begin{equation}
    L \dot{I}_L = \mathcal{G}'^{-1}(I'-I_L)- \mathcal{G}^{-1}(I_L-I) + V_n,
\end{equation}
or
\begin{equation}
    \dot{I}_L = -L^{-1}\mathcal{G}^{-1}(I_L - I) - L^{-1} \mathcal{G}'^{-1}(I_L - I') + L^{-1}V_n.
\end{equation}

\begin{equation}
\label{app:gaussian-circuit-sde}
    d I_L = -L^{-1}\mathcal{G}^{-1}(I_L - I)\, dt - L^{-1} \mathcal{G}'^{-1}(I_L - I')\, dt + L^{-1}\sqrt{S} \mathcal{N}[0, \mathbb{I}\, dt].
\end{equation}
We now proceed to non-dimensionalize the above equation. Let $\mathcal{G} = \tilde{R}^{-1} \Sigma_\pi$, $\mathcal{G}' = \tilde{R}^{-1} \Sigma_\ell$, $I_L = \tilde{I}\theta$, $I = \tilde{I} \mu_\pi$, and $I' = \tilde{I} y$. Then
\begin{equation}
    \tilde{I} d\theta = -\tilde{I}\tilde{R}L^{-1}\Sigma_\pi^{-1}(\theta - \mu_\pi)\, dt - \tilde{I}\tilde{R}L^{-1} \Sigma_\ell^{-1}(\theta - y)\, dt + L^{-1}\sqrt{S} \mathcal{N}[0, \mathbb{I}\, dt].
\end{equation}
Define $\tau = L/\tilde{R}$, giving
\begin{equation}
     d\theta = -\Sigma_\pi^{-1}(\theta - \mu_\pi)\tau^{-1}\, dt -  \Sigma_\ell^{-1}(\theta - y)\tau^{-1}\, dt + \tilde{I}^{-1}L^{-1}\sqrt{S} \mathcal{N}[0, \mathbb{I}\, dt].
\end{equation}
If we set $S = 2\tilde{I}^2 L \tilde{R}$, then we have
\begin{equation}
     d\theta = -\Sigma_\pi^{-1}(\theta - \mu_\pi)\tau^{-1}\, dt -  \Sigma_\ell^{-1}(\theta - y)\tau^{-1}\, dt + \mathcal{N}[0, 2\mathbb{I}\tau^{-1}\, dt],
\end{equation}
which is Eq. \eqref{eq:gaussian-sde}

\section{Time and Energy Cost of Gaussian-Gaussian Posterior Sampling}\label{app_TimeEnergyGG}

When the circuit in Fig. \ref{app:gaussian-circuit-fig} is operated, work is done by the voltage and current sources, energy is stored in the inductors, and heat is dissipated by the resistors. The total energy must be conserved in accordance with the first law of thermodynamics, as noted in Eq.~\eqref{eqn_thermo1stLaw},
\begin{equation}
    \label{app:first-law-thermo}
    W = \Delta E + Q,
\end{equation}
where $W$ is the work done by the voltage and current sources, $\Delta E$ is the change in internal energy of the inductors, and $Q$ is the heat dissipated by the resistors. All of these quantities are with respect to a time interval $[0,T]$, during which the circuit is allowed to evolve before a sample is taken at time $T$. At time $t=0$ we set $I_L = 0$, and so
\begin{equation}
    \Delta E = \frac{1}{2}L \left| I_L(T)\right|^2
\end{equation}
Meanwhile, each resistor dissipates heat at power $V^2/R$, so the total heat flow from the resistors is
\begin{equation}
    \dot{Q} = V^\intercal \mathcal{G} V + V'^\intercal \mathcal{G} V',
\end{equation}
or using the relations $V = \mathcal{G}^{-1}(I_L - I)$ and $V' = \mathcal{G}'^{-1}(-I_L + I')$,
\begin{equation}
   \dot{Q} = (I_L - I)^\intercal \mathcal{G}^{-1}(I_L - I) + (I_L - I')^\intercal \mathcal{G}'^{-1}(I_L - I').
\end{equation}
Note that Eq. \eqref{eq:gaussian-sde} can be written $d\theta = -\frac{1}{2} \tilde{I}^{-2}\tilde{R}^{-1}\nabla \dot{Q}\, \tau^{-1}dt +\mathcal{N}[0,2\mathbb{I}\tau^{-1}dt$, so
\begin{equation}
    d \braket{\dot{Q}} = \braket{\nabla \dot{Q} \cdot dI_L} =-\frac{1}{4} \tilde{R}^{-1}\left \langle\left|\nabla \dot{Q} \right|^2\right \rangle \tau^{-1}dt,
\end{equation}
establishing that $\braket{\dot{Q}}$ is decreasing. Completing the square, we find
\begin{equation}
    \tilde{I}^{-2}\tilde{R}^{-1}\dot{Q} = (\theta - \mu_{\theta|y}) ^\intercal \Sigma_{\theta|y}^{-1} (\theta -\mu_{\theta|y}) +\mu_\pi^\intercal \Sigma_\pi^{-1}\mu_\pi + y^\intercal \Sigma_\ell^{-1} y-\mu_{\theta|y}^{\intercal} \Sigma_{\theta|y}^{-1} \mu_{\theta|y},
\end{equation}
and because $\braket{\dot{Q}}$ is decreasing we have an upper bound
\begin{equation}
    \tilde{I}^{-2}\tilde{R}^{-1}\braket{\dot{Q}} \leq  \mu_\pi^\intercal \Sigma_\pi^{-1}\mu_\pi + y^\intercal \Sigma_\ell^{-1} y.
\end{equation}
Because $I_L(0)=0$, $\braket{\Delta E}$ is increasing. Therefore we get an upper bound by computing $\braket{\Delta E}$ in the stationary distribution
\begin{equation}
    \tilde{I}^{-2}L^{-1}\braket{\Delta E} \leq \frac{1}{2}(\text{tr}\{\Sigma_{\theta|y}\} + \mu_{\theta|y}^\intercal \mu_{\theta|y}).
\end{equation}
We now find the time needed to achieve a desired error in the squared Wasserstein distance. As the Wasserstein distance itself behaves like an absolute error rather than a relative error, we will later consider the Wasserstein distance normalized by $\|\Sigma_{\mu|\theta}\|^{1/2}$. The squared Wasserstein distance satisfies
\begin{align}
\mathcal{W}^2_t &\leq e^{-\lambda_\text{min}(\Sigma_{\theta|y}^{-1}) t/\tau}\mathcal{W}^2_0\\
&  
\leq e^{- t/\|\Sigma_{\theta|y}\|\tau}\mathcal{W}^2_0.
\end{align}
Note that for $\mu_0=0$ and $\Sigma_0=0\mathbb{I}$,
\begin{equation}
\mathcal{W}_0^2 =\text{tr}\{\Sigma_{\theta|y}\} + \mu_{\theta|y}^\intercal \mu_{\theta|y}.
\end{equation}
To achieve $\mathcal{W}_T^2 \leq \varepsilon_W^2 \|\Sigma_{\theta|y}\|$, we can set
\begin{equation}
T=\|\Sigma_{\theta|y}\|\tau \ln(\|\Sigma_{\theta|y}\|^{-1}\mathcal{W}_0^2\varepsilon_W^{-2}).
\end{equation}
We arrive at the following upper bound on the work
\begin{equation}
    W \leq \tilde{I}^2 L\|\Sigma_{\theta|y}\|(\mu_\pi^\intercal \Sigma_\pi^{-1}\mu_\pi + y^\intercal \Sigma_\ell^{-1} y) \ln(\|\Sigma_{\theta|y}\|^{-1}\mathcal{W}_0^2\varepsilon_W^{-2}) + \frac{1}{2}\tilde{I}^2 L\mathcal{W}_0^2
\end{equation}
We now must obtain an upper bound on $\mathcal{W}_0^2$. Assume that for some $\mathcal{M}_\text{max}>0$, we have $\mu_\pi^\intercal \Sigma_\pi^{-1} \mu_\pi \leq \mathcal{M}_\text{max}$ and $y^\intercal \Sigma_\ell^{-1} y \leq \mathcal{M}_\text{max}$. Note that the term $\mu_\pi^\intercal \Sigma_\pi^{-1}\mu_\pi + y^\intercal \Sigma_\ell^{-1} y-\mu_{\theta|y}^{\intercal} \Sigma_{\theta|y}^{-1} \mu_{\theta|y}$ which appeared when completing the square must be positive to preserve the positivity of the original quadratic. Therefore
\begin{align*}
\mathcal{W}_0^2 &=
\text{tr}\{\Sigma_{\theta|y}\} + \mu_{\theta|y}^\intercal \mu_{\theta|y}\\
& 
\leq \text{tr}\{\Sigma_{\theta|y}\} +  \lambda_\text{min}(\Sigma_{\theta|y}^{-1})^{-1}\mu_{\theta|y}^\intercal \Sigma_{\theta|y}^{-1} \mu_{\theta|y}  \\
& 
\leq  \|\Sigma_{\theta|y}\|(d+ 2\mathcal{M}_\text{max}).
\end{align*}
We then have
\begin{equation}
    W \leq 2\tilde{I}^2 L\|\Sigma_{\theta|y}\|\mathcal{M}_\text{max} \ln((d+ 2\mathcal{M}_\text{max})\varepsilon_W^{-2}) + \frac{1}{2}\tilde{I}^2 L\|\Sigma_{\theta|y}\|(d+ 2\mathcal{M}_\text{max})
\end{equation}
Finally we assume that $\|\Sigma_\pi\|\leq 1$ and $\|\Sigma_\ell\|\leq 1$ (which can always be assured by rescaling the original problem). In this case
\begin{equation}
    W \leq 2\tilde{I}^2 L\mathcal{M}_\text{max} \ln((d+ 2\mathcal{M}_\text{max})\varepsilon_W^{-2}) + \frac{1}{2}\tilde{I}^2 L(d+ 2\mathcal{M}_\text{max})
\end{equation}
Using these assumptions, we also get an upper bound on time
\begin{equation}
    T \leq \tau \ln ((d+2\mathcal{M}_\text{max})\varepsilon_W^{-2}).
\end{equation}

\section{Analysis of Bayesian Logistic Regression Circuit}
\label{app_CircuitLR}

We now analyze the circuit in Figure \ref{app-logistic-regression-circuit-fig}. The boxes labeled Diff. Pair represent differential pairs of NPN bipolar junction transistors (BJTs), as shown in Fig. \ref{app-diff-pair-fig}. To achieve a working implementation, additional circuitry is needed to support the differential pair and assure that it is appropriately biased, including a power source and possibly current mirrors.

The following conventions for current flow will be used
\begin{itemize}
    \item $I_C$ is the current \emph{into} the collector of a transistor. $I_B$ is the current \emph{into} the base of a transistor. $I_E$ is the current \emph{out of} the emitter of a transistor.
    \item The output current $I_o$ of a differential pair is the current that flows \emph{into} the collector of the BJT labeled $Q_a$.
    \item Positive current flows in the direction of the arrow through all current sources.
    \item Positive current flows downwards through $C_1$ and $C_2$ and from left to right through $R_{12}$.
    \item Through resistors $R_{A11}$, $R_{B11}$, etc. positive current always flows towards the base of the transistor.
    
\end{itemize}

\subsection{Analysis of the BJT differential pair}

\begin{figure}[h]
  \centering
  \includegraphics[width=0.5\linewidth]{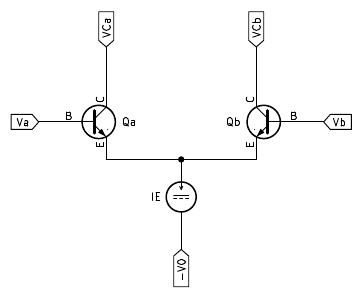}
  \caption{Circuit schematic for the BJT differential pair.}
  \label{app-diff-pair-fig}
\end{figure}
We first consider the behavior of differential pair subcircuit, which can be explained using the Ebers-Moll model. The Ebers-Moll model describes the BJT in active mode, meaning when $V_E < V_B < V_C$, and the circuit must be appropriately biased at all times to ensure the device is always in active mode. According to this model, in active mode the following relations are satisfied
\begin{equation}
\label{app:ebers-moll-eq-1}
    I_C = I_S \left(e^{(V_B - V_E)/V_T}-1\right),
\end{equation}
\begin{equation}
\label{app:ebers-moll-eq-2}
    I_C = \alpha I_E,
\end{equation}
where $I_S$ is the saturation current, $V_T$ the thermal voltage, and $\alpha$ is the common-base current gain. $I_S$ is typically on the order of $10^{-15}$ to $10^{-12}$ Amps, and at room temperature $V_T = 25.3\text{mV}$. The parameter $\alpha$ is between $0.98$ and $1$. It follows from Kirchoff's current law (KCL) that $I_B = (1-\alpha)I_E$. For these typical values of the parameters appearing in Eq. \eqref{app:ebers-moll-eq-1} the subtraction of unity in parentheses can safely be ignored, which we will do in what follows. In order for the Ebers-Moll model to be valid, the voltage $V_0$ should be determined such that $V_C>V_B>V_E$ for all transistors at all times, but the value of $V_0$ is otherwise unimportant.

To analyze the differential pair of transistors $Q_{a}$ and $Q_{b}$, observe that (by KCL)
\begin{equation}
    I_{Ea} + I_{Eb} = I_{E}.
\end{equation}
We must distinguish between the two base voltages $V_{a}$ and $V_{b}$, but the two emitter voltages are the same, so we write $V_{E} = V_{Ea} = V_{Eb}$. Using Eqs. \eqref{app:ebers-moll-eq-1} and \eqref{app:ebers-moll-eq-2} then,
\begin{equation}
    I_{E} = \frac{I_S}{\alpha} e^{-V_{E}/V_T}\left(e^{V_{a}/V_T} + e^{V_{b}/V_T}\right),
\end{equation}
where we have dropped the $-1$ as explained earlier. Now the emitter current $I_{Ea}$ can be written as
\begin{align}
    I_{Ea} & = \frac{I_S}{\alpha} e^{(V_{a} - V_{E})/V_T}\\
    & =  \frac{I_{E}e^{V_{a}/V_T}}{e^{V_{a}/V_T} + e^{V_{Bb}/V_T}}\\
    & = \frac{I_{E}}{1+e^{-(V_{a} - V_{b})/V_T}},
\end{align}
and similarly
\begin{equation}
    I_{Eb} = \frac{I_{E}}{1+e^{(V_{a} - V_{b})/V_T}}.
\end{equation}
Equation \eqref{app:ebers-moll-eq-2} is then used to find the collector currents
\begin{equation}
\label{app:collector-current-a-eq}
    I_{Ca} = \frac{\alpha I_{E}}{1+e^{-(V_{a} - V_{b})/V_T}},
\end{equation}
\begin{equation}
\label{app:collector-current-b-eq}
    I_{Cb} = \frac{\alpha I_{E}}{1+e^{(V_{a} - V_{b})/V_T}}.
\end{equation}

The base voltages $V_{a}$ and $V_{b}$ are still undetermined. However, we will assume the limit $\alpha \to 1$, where the base current goes to zero. In this limit, the two transistor bases may be connected to nodes in an external circuit to set their voltages. As there is no base current, these connections do not affect the voltages in the external circuit. In what follows, we will consider $I_{Ca}$ the output of the differential pair, and label this current $I_o$. Again taking the limit $\alpha \to 1$, we have
\begin{equation}
\label{app-diff-pair-characteristic}
    I_o = \frac{ I_{E}}{1+e^{(V_{a} - V_{b})/V_T}} = I_E L(-(V_a - V_b)/V_T),
\end{equation}
where $L(z) = 1/(1+e^{-z})$ is the standard logistic function. Note that the support circuitry may include a current mirror that inverts the sign of the output current. As this formally has the same effect as a negative value of $I_E$, we will allow $I_E$ to be negative in what follows.

\subsection{Analysis of the logistic regression circuit}

\begin{figure}[h]
  \centering
  \includegraphics[width=\linewidth]{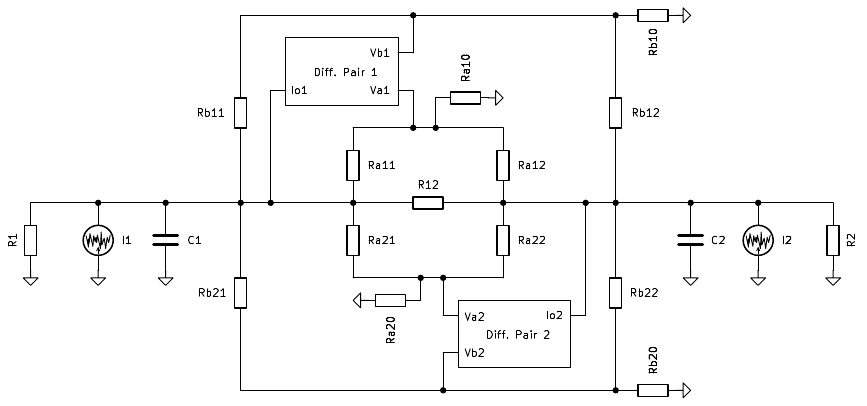}
  \caption{Circuit schematic for the logistic regression posterior sampling device.}
  \label{app-logistic-regression-circuit-fig}
\end{figure}

As the BJT bases draw negligible current, the voltages $V_{a1}$, $V_{b1}$, $V_{a2}$, and $V_{b2}$ in the circuit can be determined by considering the circuit in the absence of the differential pairs. In this case, we see that (by KCL)
\begin{equation}
    R_{a11}^{-1}(V_{C1} - V_{a1}) + R_{a12}^{-1}(V_{C2} - V_{a1}) - R_{a10}^{-1}V_{a1}=0,
\end{equation}
and solving for $V_{a1}$ gives
\begin{equation}
    V_{a1} = \frac{R_{a11}V_{C1} + R_{a12} V_{C2}}{R_{a11}  + R_{a12} + R_{a11}R_{a12}R_{a10}^{-1}} =
\frac{R_{a11}^{-1} V_{C1} + R_{a12}^{-1} V_{C2}}{ R_{a10}^{-1}+R_{a11}^{-1} + R_{a12}^{-1} }.
\end{equation}
The same reasoning applies for $V_{b1}$, resulting in
\begin{equation}
   V_{b1} = \frac{R_{b11}^{-1} V_{C1} + R_{b12}^{-1} V_{C2}}{ R_{b10}^{-1}+R_{b11}^{-1} + R_{b12}^{-1}},
\end{equation}
so
\begin{equation}
    V_{a1} - V_{b1} = \frac{g_{a11} V_{C1} + g_{a12} V_{C2}}{g_{a10} +g_{a11} + g_{a12}} - \frac{g_{b11} V_{C1} + g_{b12} V_{C2}}{g_{b10} + g_{b11} + g_{b12} } ,
\end{equation}
where we have written the previous results in terms of the conductance $g = R^{-1}$. The above can be written more conveniently by defining the vectors $\hat{g}_{a1} = (g_{a10} + g_{a11} + g_{a12})^{-1}(g_{a11},  g_{a12})^\intercal$ and $\hat{g}_{b1} = (g_{b10} + g_{b11} + g_{b12})^{-1}(g_{b11},  g_{b12})^\intercal$, in terms of which we have
\begin{equation}
    V_{a1} - V_{b1} = (\hat{g}_a - \hat{g}_b)^\intercal V_C.
\end{equation}
or, defining $\hat{g}_1 = \hat{g}_{a1} - \hat{g}_{b1}$, we simply have
\begin{equation}
     V_{a1} - V_{b1} = \hat{g}_1^\intercal V_C.
\end{equation}
The latter result can be plugged into Eq. \eqref{app-diff-pair-characteristic} to get $I_{o1}$,
\begin{equation}
    I_{o1} = I_{E1} L(-\hat{g}_1^\intercal V_C/V_T),
\end{equation}
where, as before, $L(z) = 1/(1+e^{-z})$ is the standard logistic function. By an identical derivation to the one above, a similar relation holds for the lower subcircuit
\begin{equation}
    I_{o2} = I_{E2} L(-\hat{g}_2^\intercal V_C/V_T).
\end{equation}
We also assume that all resistors $R_{aij}$, $R_{bij}$ are very large compared to $R_{12}$ so the current flowing through these resistors can be treated as negligible. This assumption does not affect the function of resistors $R_{aij}$, $R_{bij}$ because only the ratios of these resistances determine the voltages $V_{ai}$, $V_{bi}$.
Next, we apply KCL to the nodes at the top of capacitors $C_1$ and $C_2$
\begin{equation}
   -I_{C1} +  I_1 - R_1^{-1}V_{C1} + R_{12}^{-1}(V_{C2} - V_{C1})-I_{o1}=0,
\end{equation}
Similarly, KCL for the node above capacitor $C_2$ reads
\begin{equation}
   -I_{C2} +  I_2 -  R_2^{-1}V_{C2} + R_{12}^{-1}(V_{C1} - V_{C2})-I_{o2}=0.
\end{equation}
Substituting in the expressions derived for the collector currents, we then have
\begin{equation}
   -I_{C1} +  I_1 - R_1^{-1}V_{C1} + R_{12}^{-1}(V_{C2} - V_{C1})-I_{E1} L(-\hat{g}_1^\intercal V_C/V_T)=0,
\end{equation}
\begin{equation}
   -I_{C2} +  I_2 - R_2^{-1} V_{C2} + R_{12}^{-1}(V_{C1} - V_{C2})-I_{E2} L(-\hat{g}_2^\intercal V_C/V_T)=0.
\end{equation}
Next we define the conductance matrix
\begin{equation}
    \mathcal{G} = \begin{pmatrix}
R_1^{-1} + R_{12}^{-1} & - R_{12}^{-1} \\
-R_{12}^{-1} & R_{2}^{-1} + R_{12}^{-1},
    \end{pmatrix},
\end{equation}
allowing us to write a single vector equation
\begin{equation}
-I_C + I - \mathcal{G}V_C - I_E L(-\hat{g}^\intercal V_C/V_T) = 0, 
\end{equation}
where we have also set $\hat{g}_1 = \hat{g}_2$. Now using the fact that $dV_C/dt = C^{-1}I_C$, we have the following vector differential equation
\begin{equation}
    C\frac{d V_C}{dt}   =  - \mathcal{G}V_C - I_E L(-\hat{g}^\intercal V_C/V_T)  + I.
\end{equation}
We assume the current vector $I$ has a DC component $I_{DC}$ and a noise component $I_\text{noise}$. The noise component is assumed to be an ideal white noise process of infinite bandwidth and power spectral density $S$, which we write $I_\text{noise} = \sqrt{S}\xi(t)$. Altogether, we get the stochastic differential equation
\begin{equation}
    dV_C = -C^{-1} \mathcal{G}V_C\, dt + C^{-1}I_{DC}\, dt - C^{-1} \alpha I_EL(\hat{g}^\intercal V_C/V_T)\, dt + C^{-1}\sqrt{S} \xi(t) \, dt.
\end{equation}
Using the identity $\xi(t)\,dt = \mathcal{N}[0,dt]$, this becomes
\begin{equation}
    dV_C = -C^{-1} \mathcal{G}V_C\, dt + C^{-1}I_{DC}\, dt - C^{-1} \alpha I_EL(\hat{g}^\intercal V_C/V_T)\, dt + C^{-1}\sqrt{S} \mathcal{N}[0, dt].
\end{equation}
At this point it is convenient to define dimensionless quantities which are mapped to the physical parameters of the circuit. Define $\theta = V_C/\tilde{V}$, $\Sigma^{-1} = \tilde{R}\mathcal{G}$, $\Sigma^{-1}\mu = I_{DC}/\tilde{I}$, and $yx=-I_E/\tilde{I}$. Our equation now takes the form
\begin{equation}
    \tilde{V} dx = -\tilde{V}\tilde{R}^{-1}C^{-1} \Sigma^{-1}x\, dt + C^{-1}\tilde{I}\Sigma^{-1}\mu\, dt + C^{-1}  \tilde{I} L(-\hat{g}^\intercal V_C/V_T)yx\, dt + C^{-1}\sqrt{S} \mathcal{N}[0, \mathbb{I}\,dt].
\end{equation}
Next, let $\tau = \tilde{R} C$, and set $\tilde{I}  =\tilde{V} C/\tau$ and $S = 2\tilde{V}^2 C^2/\tau$. In this case,
\begin{equation}
     d\theta = - \Sigma^{-1}\theta \tau^{-1} dt + \Sigma^{-1}\mu\tau^{-1} dt +  L(-\hat{g}^\intercal V_C/V_T)yx\tau^{-1} dt +  \mathcal{N}[0, 2\mathbb{I}\tau^{-1}dt].
\end{equation}
Finally, we set $\hat{g} = yxV_T/\tilde{V}$ to obtain
\begin{equation}
     d\theta = -\Sigma^{-1}(\theta-\mu) \tau^{-1}dt +   L(-y\theta^\intercal x )yx\tau^{-1}dt +  \mathcal{N}[0, 2\mathbb{I}\tau^{-1}dt],
\end{equation}
which is identical to Eq. \eqref{eq:logistic-langevin-eq}.

\section{Time Cost of Logistic Regression}\label{app_TimeLR}
We again use the fact that when $-\nabla^2 \ln p_{\theta|y} \geq \alpha \mathbb{I}$ everywhere, we have
\begin{equation}
    \mathcal{W}^2_t \leq e^{\alpha t/\tau}\mathcal{W}^2_0.
\end{equation}
For logistic regression we have
\begin{equation}
    -\ln p_{\theta|y} = \frac{1}{2} (\theta - \mu_\pi)^\intercal \Sigma_\pi^{-1} (\theta - \mu_\pi) - \ln (L(y\theta^\intercal x)).
\end{equation}
Taking the Hessian of the second term gives
\begin{equation}
  -  \partial_{\theta_i}\partial_{\theta_j}  \ln(L(y \theta^\intercal x)) = \frac{e^{y \theta^\intercal x}}{(e^{y \theta^\intercal x}+1)^2}x_i x_j,
\end{equation}
using the fact that $y=\pm 1$ so $y^2 = 1$. Therefore the Hessian is
\begin{equation}
    -\nabla^2 \ln (L(y\theta^\intercal x)) = 
\frac{e^{y \theta^\intercal x}}{(e^{y \theta^\intercal x}+1)^2}x x^\intercal.
\end{equation}
Obviously all of the eigenvalues of the above matrix are non-negative, so overall we have
\begin{equation}
    -\nabla^2 \ln p_{\theta|y} \geq \lambda_\text{min}(\Sigma_\pi^{-1})\mathbb{I}.
\end{equation}
This results on the following bound on the Wasserstein distance to the posterior
\begin{align}
\mathcal{W}^2_t &\leq e^{-\lambda_\text{min}(\Sigma_\pi^{-1}) t/\tau}\mathcal{W}^2_0\\
&  
\leq e^{- t/\|\Sigma_\pi\|\tau}\mathcal{W}^2_0.
\end{align}
To achieve $\mathcal{W}_T^2 \leq \varepsilon_W^2 \|\Sigma_{\theta|y}\|$, we can set
\begin{equation}
T=\|\Sigma_\pi\|\tau \ln(\|\Sigma_{\theta|y}\|^{-1}\mathcal{W}_0^2\varepsilon_W^{-2}).
\end{equation}
The next task is to identify an upper bound on $\|\Sigma_{\theta|y}\|^{-1}\mathcal{W}_0^2$ and . Recall that the definition of the Wasserstein 2 distance is a minimization problem over the set of joint distributions having the correct marginals
\begin{equation}
    \mathcal{W} (\rho_1, \rho_2) = \inf (\mathbb{E}|X_1 - X_2|^2)^{1/2},
\end{equation}
and so any joint distribution with the correct marginals can be used to give an upper bound. In particular, we consider the case where $X$ and $Y$ are independent, with $X_1=0$ and $X_2\sim p_{\theta|y}$, and find that
\begin{align}
\mathcal{W}_0^2 &\leq \braket{X_2^\intercal X_2}\\
& = \text{tr}\{\Sigma_{\theta|y}\} + \mu_{\theta|y}^\intercal \mu_{\theta|y}\\
&\leq \text{tr}\{\Sigma_{\theta|y}\} +  \lambda_\text{min}(\Sigma_{\theta|y}^{-1})^{-1}\mu_{\theta|y}^\intercal \Sigma_{\theta|y}^{-1} \mu_{\theta|y}  \\
& 
\leq  \|\Sigma_{\theta|y}\|(d+ \mathcal{M}_{\theta|y}),
\end{align}
where $\mathcal{M}_{\theta|y} = \mu^\intercal_{\theta|y} \Sigma_{\theta|y}^{-1} \mu_{\theta|y}$. Unfortunately it is less clear how to bound $\mathcal{M}_{\theta|y}$ in terms of the prior and likelihood parameters, and we leave this task for future work. At any rate, we arrive at
\begin{equation}
    T \leq \|\Sigma_\pi\|\tau \ln((d+ \mathcal{M}_{\theta|y})\varepsilon_W^{-2}).
\end{equation}
As in the case of the Gaussian-Gaussian model, without loss of generality we may again assume $\|\Sigma_\pi\| \leq 1$ to obtain
\begin{equation}
    T \leq\tau \ln((d+ \mathcal{M}_{\theta|y})\varepsilon_W^{-2}).
\end{equation}

\section{Conditioning on multiple I.I.D. samples}\label{app_MultipleSamples}
When conditioning on a single sample $y$, the energy $U$ can be separated into two terms, one mapping to the prior and the other to the likelihood:
\begin{equation}
    U(r) =  U_\pi(r) +  U_\ell(r),
\end{equation}
where $\beta U_\pi(r) = -\ln p_\theta(r/\tilde{r})$ and $\beta U_\ell(r) = -\ln p_{y|\theta}(y|r/\tilde{r})$. In general we may have a number of I.I.D. samples $Y=(y_1, \dots y_N)$, and would like to sample from $p_{\theta|Y}(\theta|Y)$. Because the samples of $y$ are I.I.D., we have
\begin{equation}
    p_{Y|\theta}(Y|\theta) = \prod_{i=1}^N p_{y|\theta}(y_i|\theta).
\end{equation}
In this case the likelihood part of the potential energy takes the form
\begin{equation}
    \beta U_\ell(r) = -\sum_{i=1}^N \ln p_{y|\theta}(y_i|r/\tilde{r}),
\end{equation}
while the prior part is the same as in the single-sample case, $\beta U_\pi(r) = -\ln p_\theta (r/\tilde{r})$. This form of the potential energy has a convenient physical interpretation: the function $U_\pi$ can be interpreted as the self-energy of the system in state $r$ (that is when it is decoupled from an external system), while the function $U_\ell(\theta)$ can be viewed as an interaction energy between the state $r$ and the state $y$ of an external system. When there are multiple I.I.D. samples, this is analogous to the state $r$ interacting with a collection of external systems in states $Y = ( y_1 \dots y_N )$, and each such interaction contributes its own term to the interaction energy. This provides a framework for building a physical device to sample from the posterior conditioned on multiple I.I.D. samples; one must simply couple a collection of external systems in states $Y = (y_1\dots y_N)$ to the system in such a way that each interaction contributes an energy of $-\ln p_{y|\theta}(y|r/\tilde{r})$.

We will now describe another approach to building a physical device that samples from the posterior conditioned on multiple I.I.D. samples of $y$. We first observe that the Langevin equation for the device in this case must be
\begin{equation}
    \label{eq:langevin-iid}
    d\theta= \nabla_\theta \ln p_{\theta}(\theta)\tau^{-1}\,dt +\sum_{i=1}^N \nabla_\theta \ln p_{y|\theta}(y_i|\theta)\, dt+ \mathcal{N}[0, 2  \tau^{-1}dt],
\end{equation}
As discussed above, if we have a device that can implement the $N$ likelihood drift terms simultaneously then the problem his solved. However, suppose that we have a device that is only capable of implementing a single likelihood term at a time, but $y$ may be varied as a function of time. Additionally, we make the interaction energy for this device larger by a factor of $N$ for reasons that will become clear. That is, we have a device that implements an SDE of the form
\begin{equation}
    d\theta= \nabla_\theta \ln p_{\theta}(\theta)\tau^{-1}\,dt + N\nabla_\theta \ln p_{y|\theta}(y(t)|\theta)\, dt+ \mathcal{N}[0, 2  \tau^{-1}dt].
\end{equation}
We may choose a short time duration $\Delta t$, and set \begin{equation}
    y(t) = y_{\lfloor t/\Delta t \rfloor \text{  mod  } N + 1}.
\end{equation}
So for $0\leq  t \leq \Delta t$ we set $y(t) = y_1$, for $\Delta t < t \leq 2 \Delta t$ we set $y(t) = y_2$, and so on. Once $t > N \Delta t$ we start over at $y_1$ and continue cycling over all of the I.I.D. samples. Suppose that $\Delta t$ is short enough that all of the samples are cycled over before the state $\theta$ changes significantly. We may then average drift term $N \nabla_\theta \ln p_{y|\theta}$ over a period of time $N\Delta t$ and consider $\theta$ constant within this average. Carrying out this time average, we find
\begin{equation}
     \frac{1}{N \Delta t}\sum_{i=1}^N \Delta t N \nabla_\theta \ln p_{y|\theta}(y_i|\theta)
= \sum_{i=1}^N \nabla_\theta p_{y|\theta}(y_i|\theta),
\end{equation}
resulting in the correct form of the Langevin equation.

\end{document}